\documentclass[showpacs,twocolumn,aps,prl]{revtex4}
\usepackage{amsmath,amssymb}
\begin{document}
\def\cs#1#2{#1_{\!{}_#2}}
\def\css#1#2#3{#1^{#2}_{\!{}_#3}}
\def\ket#1{|#1\rangle}
\def\bra#1{\langle#1|}
\def\dbl{\hbox{${1\hskip -2.4pt{\rm l}}$}}
\def\bfh#1{\bf{\hat#1}}

\title{Disproof of Bell's Theorem by Clifford Algebra Valued Local Variables}

\author{Joy Christian}

\email{joy.christian@wolfson.ox.ac.uk}

\affiliation{Perimeter Institute, 31 Caroline Street North, Waterloo, Ontario N2L 2Y5, Canada,}
\affiliation{Department of Physics, University of Oxford, Parks Road, Oxford OX1 3PU, England}

\begin{abstract}
It is shown that Bell's theorem fails for the Clifford algebra valued local realistic
variables. This is made evident by exactly reproducing quantum mechanical expectation
value for the EPR-Bohm type spin correlations observable by means of a local, deterministic,
Clifford algebra valued variable, without necessitating either remote contextuality or
backward causation. Since Clifford product of multivector variables is non-commutative in
general, the spin correlations derived within our locally causal model violate the CHSH
inequality just as strongly as their quantum mechanical counterparts.
\end{abstract}


\maketitle

Unlike our basic theories of space and time, quantum mechanics is not a locally causal theory
\cite{Bell-La}. This fact was famously brought forth by Einstein, Podolsky, and Rosen (EPR) in
1935 \cite{EPR}. They hoped, however, that perhaps one day quantum mechanics may be
``completed'' into a realistic, locally causal theory, say  by appending the incomplete
quantum mechanical description of physical reality with additional ``hidden'' parameters. Today
such hopes of restoring locality in physics while maintaining realism seem to have been severely
undermined by Bell's theorem and its variants \cite{Bell-1964}\cite{GHSZ}, with substantive
support from experiments \cite{Clauser}\cite{Shimony-2004}. These theorems set out to prove
that {\it no physical theory which is realistic as well as local in a specified sense can
reproduce all of the statistical predictions of quantum mechanics} 
\cite{Bell-1964}\cite{Shimony-2004}. The purpose of this letter is to question the legitimacy
of this conclusion by first constructing an exact local realistic model for the EPR-Bell type
spin correlations, and then reevaluating Bell's proof of his theorem in the light of this model.
In particular, it will be shown that the much studied CHSH inequality in this context
\cite{Clauser} is violated within our local model and extended to the extrema of
${\pm\,2\sqrt{2}}$, in {\it exactly} the same manner as it is within quantum mechanics.

To prove his theorem Bell employed a rather simple argument. His goal was to show that at
least some of the predictions of quantum mechanics cannot be mimicked by a locally causal
theory. Based on Bohm's spin version of the EPR thought experiment \cite{Bohm-1951}, he
considered a pair of spin one-half particles, moving freely after production in opposite
directions, with particles ${1}$ and ${2}$ subject, respectively, to spin measurements
along independently chosen unit directions ${\bf a}$ and ${\bf b}$, which can be located at
a spacelike distance from each other. If initially the pair has vanishing total spin,
then its quantum mechanical spin state would be the entangled singlet state
\begin{equation}
|\Psi_{\bf n}\rangle=\frac{1}{\sqrt{2}}\Bigl\{|{\bf n},\,+\rangle_1\otimes
|{\bf n},\,-\rangle_2\,-\,|{\bf n},\,-\rangle_1\otimes|{\bf n},\,+\rangle_2\Bigr\}\,,
\label{single}
\end{equation}
with ${\bf n}$ indicating an arbitrary unit direction, and
\begin{equation}
{\boldsymbol\sigma}\cdot{\bf n}\,|{\bf n},\,\pm\rangle\,=\,
\pm\,|{\bf n},\,\pm\rangle \label{spin}
\end{equation}
describing the quantum mechanical eigenstates in which the particles have spin ``up'' or
``down'' in units of ${\hbar=2}$. Here ${\boldsymbol\sigma}$ is the familiar Pauli spin
``vector'' ${({\sigma_x},\,{\sigma_y},\,{\sigma_z})}$. Our interest lies in comparing the
quantum predictions of spin correlations between the two remote subsystems with
those derived from any locally causal theory.

Now, quantum mechanically the rotational invariance of the state ${\ket{\Psi_{\bf n}}}$
ensures that the expectation values of the individual spin observables
${{\boldsymbol\sigma}_1\cdot{\bf a}}$ and ${{\boldsymbol\sigma}_2\cdot{\bf b}}$ are
\begin{align}
{\cal E}_{q.m.}({\bf a})\,=\,\langle\Psi_{\bf n}&|\,{\boldsymbol\sigma}_1
\cdot{\bf a}\otimes\dbl\,|\Psi_{\bf n}\rangle\,=\,0\,\;\;{\rm and} \notag \\
{\cal E}_{q.m.}({\bf b})\,=\,\langle\Psi_{\bf n}&|\,\dbl\otimes{\boldsymbol\sigma}_2
\cdot{\bf b}\,|\Psi_{\bf n}\rangle\,=\,0\,,\label{singof}
\end{align}
where ${\dbl}$ is the identity matrix. The expectation value of the joint observable
${{\boldsymbol\sigma}_1\cdot{\bf a}\otimes{\boldsymbol\sigma}_2\cdot{\bf b}}$, on the
other hand, is
\begin{equation}
{\cal E}_{q.m.}({\bf a},\,{\bf b})\,=\,
\langle\Psi_{\bf n}|\,{\boldsymbol\sigma}_1\cdot{\bf a}\,\otimes\,
{\boldsymbol\sigma}_2\cdot{\bf b}\,|\Psi_{\bf n}\rangle\,=\,-\,
{\bf a}\cdot{\bf b}\,,\label{twoobserve}
\end{equation}
regardless of the relative distance between the two remote locations represented by the unit
vectors ${\bf a}$ and ${\bf b}$. The last result can be derived \cite{Peres-1993}
using the well known identity
\begin{equation}
(i\,{\boldsymbol\sigma}\cdot{\bf a})\,(i\,{\boldsymbol\sigma}\cdot{\bf b})\,=\,-
\,{\bf a}\cdot{\bf b}\,\dbl\,-\,
i\,{\boldsymbol\sigma}\cdot({\bf a}\times{\bf b})\,,\label{nowobserve}
\end{equation}
which follows from the non-commutativity of products of the Pauli matrices ${\sigma_j}$
(${j=x,y,z}$) defined by the algebra
\begin{equation}
\sigma_j\sigma_k=\,\delta_{jk}\,{\dbl}\,+\,i\,\epsilon_{jkl}\,\sigma_l\,,\label{paulialgebra}
\end{equation}
where ${\delta_{jk}}$ is the Kronecker delta, ${i\equiv\sqrt{-1}}$ is the unit imaginary,
and ${\epsilon_{jkl}}$ is the Levi-Civita alternating symbol.

Suppose now we consider a {\it complete} specification of the physical state of our two-state
system denoted by ${\lambda}$, specifying all of the elements of physical reality of the pair
at a suitable instant, in the manner envisaged by EPR \cite{EPR}. Here the complete state
${\lambda}$ can be taken to be discrete or continuous, a single variable or a set of variables,
a single function or a set of functions, and can even govern the measurement outcomes of spin
stochastically rather than deterministically. For our purposes, however, it would suffice to
take ${\lambda}$ as a single, continuous variable, fully compatible with deterministic laws of
motion. If we now denote by ${\rho(\lambda)}$ the normalized probability measure on the space
${\Lambda}$ of complete states, then the expectation value of the product of the two outcomes
of spin measurements, parameterized by ${\bf a}$ and ${\bf b}$ as before, can be written as
\begin{equation}
{\cal E}_{h.v.}({\bf a},\,{\bf b})\,=\int_{\Lambda}
A_{\bf a}(\lambda)\,B_{\bf b}(\lambda)\;\,d\rho(\lambda)\,,\label{prob}
\end{equation}
where ${A_{\bf a}(\lambda)}$ and ${B_{\bf b}(\lambda)}$, with values ${\pm 1}$, represent
the possible outcomes of measurements on the subsystems 1 and 2, satisfying the perfect
correlations constraint ${{\cal E}_{h.v.}({\bf n},\,{\bf n}) = -\,1}$ adapted by EPR.
What is crucial to note here is that the function ${A_{\bf a}(\lambda)}$ does not depend
on the remote context ${\bf b}$, and likewise the function ${B_{\bf b}(\lambda)}$ does not
depend on the remote context ${\bf a}$, thus adhering to the vital condition of locality
assumed by EPR. It is important to note also that the probability measure ${\rho(\lambda)}$
in (\ref{prob}) is required to depend only on ${\lambda}$, and not on either ${\bf a}$ or
${\bf b}$, which are thereby rendered ``freely chosen'' detector settings at a later time
\cite{Bell-La}. Moreover, the factorized form (\ref{prob}) of the joint expectation value of the
two outcomes can be derived explicitly as a {\it consequence} of the relativistic notion of
local causality \cite{Bell-La}. As a result, Bell's theorem reduces simply to a claim of
impossibility to reproduce the quantum mechanical correlations (\ref{twoobserve}), by means
of a local realistic expectation value of the form (\ref{prob}).

Before formally proving the theorem, however, Bell provides an illustration of the tension
between quantum mechanics and local causality by means of a local model. The idea behind the
model is to attempt to reproduce the quantum mechanical correlations (\ref{twoobserve}) for a
pair of spin one-half particles with zero total spin, and argue that it cannot be done without
admitting remote contextuality. The space ${\Lambda}$ of complete states for the model consists
of unit vectors ${\boldsymbol{\lambda}}$ in three-dimensional Euclidean space
${{\mathbb E}_3}$, with dynamical variables ${A_{\bf a}({\boldsymbol\lambda})}$ and
${B_{\bf b}({\boldsymbol\lambda})}$ defined by
\begin{equation}
A_{\bf n}({\boldsymbol\lambda})\,=\,-\,
B_{\bf n}({\boldsymbol\lambda})\,=\,{sign}\,({\boldsymbol\lambda}\cdot{\bf n})\,,
\label{bell-c1}
\end{equation}
provided ${{\boldsymbol\lambda}\cdot{\bf n}\,\not=\,0}$, and otherwise equal to the sign
of the first nonzero term from the set ${\{n_x,\,n_y,\,n_z\}}$. This simply means that
${A_{\bf n}({\boldsymbol\lambda})=+\,1}$ if the two unit vectors ${\bf n}$ and
${\boldsymbol\lambda}$ happen to point through the same hemisphere centered at the origin
of ${\boldsymbol{\lambda}}$, 
and ${A_{\bf n}({\boldsymbol\lambda})=-\,1}$ otherwise. As a visual aid
to Bell's model \cite{Peres-1993} one can think of a bomb at rest exploding into two
freely moving fragments with angular momenta ${{\boldsymbol\lambda}={\bf J}_1=-{\bf J}_2}$,
with ${{\bf J}_1+{\bf J}_2=0}$. The two outcomes ${A_{\bf a}({\bf J}_1)}$ and
${B_{\bf b}({\bf J}_2)}$ can then be taken as ${{sign}\,({\boldsymbol\lambda}\cdot{\bf a})}$
and ${{sign}\,(\,-\,{\boldsymbol\lambda}\cdot{\bf b})}$, respectively. If the initial
directions of the two angular momenta are uncontrollable but describable by an isotropic
probability distribution ${\rho({\boldsymbol\lambda})}$, then the local realistic expectation
values of the individual outcomes can be easily worked out to be \cite{Peres-1993}
\begin{equation}
{\cal E}_{h.v.}({\bf n})\,=\,\pm\int_{{\mathbb E}_3}
{sign}\,({\boldsymbol\lambda}\cdot{\bf n})\;\,d\rho({\boldsymbol\lambda})\,=\,0\,,
\label{class}
\end{equation}
where ${{\bf n}={\bf a}}$ or ${\bf b}$; and their joint expectation value based
on the local form (\ref{prob}) can be similarly worked out to be
\begin{equation}
{\cal E}_{h.v.}({\bf a},\,{\bf b})\,=\,-1+\frac{2}{\pi}\,
\cos^{-1}\left({\bf a}\cdot{\bf b}\right).\label{classprob}
\end{equation}
Comparing this local realistic correlation function with its quantum mechanical counterpart
(\ref{twoobserve}), it is frequently stressed in the established literature \cite{Peres-1993}
that
\begin{equation}
\left|\,{\cal E}_{q.m.}({\bf a},\,{\bf b})\right|\,
\geqslant\,
\left|\,{\cal E}_{h.v.}({\bf a},\,{\bf b})\right|.
\label{dynamical}
\end{equation}
Thus quantum correlations are claimed to be stronger than any local realistic possibility, in
almost all but the cases where both are either ${0}$ or ${\pm 1}$. Indeed, it is claimed that
``quantum phenomena are more disciplined'' than their ``classical'' counterparts
\cite{Peres-1993}. What is more, Bell has surmised \cite{Bell-1964} that local realistic
correlations such as (\ref{classprob}) cannot be amended to recover the quantum correlations
(\ref{twoobserve}), without necessitating remote contextuality \cite{Shimony-2004}.

It is at this stage that our skepticism of Bell's theorem hardens. Although, as yet, we
are only at the heuristic stage of the formal proof of his theorem, conceptually it is an
important stage, and before reconsidering his proof it is worth investigating whether
more realistic models for spin can perhaps replace the inequality (\ref{dynamical}) with
an {\it exact} equality. After all, spin angular momentum within classical physics is
usually represented, not by a polar vector, but by an axial or pseudo vector, composed
of a cross product of two polar vectors. Moreover, as we saw above, within quantum
mechanics the physics of spin one-half particles is intimately linked to the Pauli
algebra (\ref{paulialgebra}). Could then incorporating such realistic features into
Bell's local model for spin make any difference? As we shall see, even a minimum of such
amendments to Bell's model has devastating consequences for his theorem.

To appreciate this assertion, let us first recall that Pauli matrices ${\{\sigma_j\}}$
generating the algebra (\ref{paulialgebra}) actually form a matrix representation of
the Clifford algebra ${{Cl}_{3,0}}$ of the orthogonal directions in the Euclidean space
${{\mathbb E}_3}$
\cite{Clifford}. It is then hardly surprising that Clifford algebra ${{Cl}_{3,0}}$
can be generated also by the set of orthonormal basis vectors ${\{\,{\bf e}_j\}}$ of the
physical vector space ${{\mathbb E}_3}$, defined by
\begin{equation}
{\bf e}_j\,{\bf e}_k\,:=\,\delta_{jk}\,+\,I\;\epsilon_{jkl}\;{\bf e}_l\,
\equiv\,{\bf e}_j\cdot{\bf e}_k+\,
{\bf e}_j\wedge\,{\bf e}_k\,,\label{c-algebra}
\end{equation}
where ``${\,\cdot\,}$'' and ``${\wedge}$'' denote the inner and outer products, and
${j=x,\,y,}$ or ${z}$. The formal similarities between the relations (\ref{paulialgebra})
and (\ref{c-algebra}) should not be allowed to obscure the profound differences between
them. It is crucial to note that the ${{\bf e}_j}$ appearing in the above definition are not
the usual self-adjoint operators on a complex Hilbert space, but are the ordinary 3-vectors
in the real physical vector space. Moreover, the ${I=\sqrt{-1}}$ appearing
therein is not the unit imaginary ${i=\sqrt{-1}}$, but a {\it real} geometric entity
defined by ${I:={{\bf e}_x\,{\bf e}_y\,{\bf e}_z}}$, with
${\{\,{\bf e}_x,\,{\bf e}_y,\,{\bf e}_z\}}$ being a choice of right-handed frame of
orthonormal vectors, and is known variously as a pseudoscalar,
a volume form, or a directed volume element \cite{Clifford}. In fact, the algebra of the
physical space is spanned by this trivector ${I}$, along with a scalar, the vectors
${\{\,{\bf e}_j\}}$, and the bivectors ${\{\,{\bf e}_j\wedge\,{\bf e}_k\}}$. The Euclidean
vector space ${{\mathbb E}_3}$ is then defined simply as a set of all vectors ${\bf x}$
satisfying the equation ${I\wedge{\bf x}=\,0}$, with the basic product between its elements
defined as
\vspace{-0.2cm}
\begin{equation}
{\bf x}\;{\boldsymbol\xi}\,:=\,{\bf x}\cdot{\boldsymbol\xi}\,+\,{\bf x}\wedge
{\boldsymbol\xi}\,,
\label{fun-cliff}
\end{equation}
where ${\boldsymbol\xi}$ is any homogeneous multivector in ${{Cl}_{3,0}}$.
Thus, by subsuming it as a subsystem, Clifford algebra is said to have
``completed'' the vector algebra of Gibbs \cite{Clifford}.

Returning to Bell's local model, we begin by observing that within Clifford algebra a
rotation of a physical object is represented by a {\it bivector}, which can be visualized
as an {\it oriented} parallelogram, composed of two vectors \cite{Clifford}. Taking aboard
this hint, let us then venture to replace the polar vector ${\boldsymbol\lambda}$ of Bell's
model with the unit trivector
\vspace{-0.4cm}
\begin{equation}
{\boldsymbol\mu}\,=\,{\bf u}\,\wedge\,{\bf v}\,\wedge\,{\bf w}\,=\,\pm\,I\,\equiv\,
\pm\;{{\bf e}_x}\,\wedge\,{{\bf e}_y}\,\wedge\,{{\bf e}_z}\,,\label{tri-mu}
\end{equation}
which can be pictured as a parallelepiped of unit volume, assembled by the vectors
${\bf u}$, ${\bf v}$, and ${\bf w}$ of finite lengths and arbitrary directions, giving it
an unspecified shape and orientation. Here the second of the equalities follows from the
fact that every trivector in the algebra ${{Cl}_{3,0}}$ differs from ${I}$ only by its
volume and orientation. This allows us to quantify the ambivalence in the orientation
of ${\boldsymbol\mu}$ simply by the sign of ${I}$. In what follows, we shall take the
vectors ${\bf u}$, ${\bf v}$, and ${\bf w}$---and hence ${\boldsymbol\mu}$---to be
uncontrollable, but describable by an isotropic probability distribution
${{\boldsymbol\rho}({\boldsymbol\mu})}$. Thus, in essence, the intrinsic freedom of
choice in the {\it initial} orientation of the unit pseudoscalar ${\boldsymbol\mu}$
would be our ``local hidden variable.'' The local analogue of the spin variable
${{\boldsymbol\sigma}\cdot{\bf n}}$ can then be taken as the projection
${{\boldsymbol\mu}\cdot{\bf n}}$ ${=}$ ${\pm\,I\,{\bf n}\,}$, which turns out to
be a {\it unit} bivector
\vspace{-0.2cm}
\begin{equation}
{\boldsymbol\mu}\cdot{\bf n}\,\equiv\,\pm\,
\{\,n_x\;{{\bf e}_y}\,\wedge\,{{\bf e}_z}
\,+\,n_y\;{{\bf e}_z}\,\wedge\,{{\bf e}_x}
\,+\,n_z\;{{\bf e}_x}\,\wedge\,{{\bf e}_y}\}.
\label{mu}
\end{equation}
Since ordinary vectors in ${{\mathbb E}_3}$, such as the unit vectors ${\bf n}$ [being
solutions of ${{\boldsymbol\mu}\wedge{\bf x}}$ ${=}$ ${(\pm\,I)\wedge{\bf x}}$ ${=}$
${\pm\,(I\wedge{\bf x})=0}$], are {\it insensitive} to the sign ambiguity in ${\boldsymbol\mu}$,
the bivector (\ref{mu}) provides us a natural pair of dichotomic observables,
\vspace{-0.4cm}
\begin{equation}
A_{\bf n}({\boldsymbol\mu})=
B_{\bf n}({\boldsymbol\mu})={\boldsymbol\mu}\cdot{\bf n}
\cong\pm\,1\in S^2\;{\rm about}\;{\bf n}\in{\mathbb E}_3\,,
\label{bell-c2}
\end{equation}
reflecting isomorphism between ${S^2}$ and the space of
${{\boldsymbol\mu}\cdot{\bf n}}$.
Moreover, Clifford product of two such bivectors gives
\vspace{-0.1cm}
\begin{align}
(\,{\boldsymbol\mu}\cdot{\bf a})(\,{\boldsymbol\mu}\cdot{\bf b})
&=(\,{\boldsymbol\mu}\,{\bf a}\,)(\,{\boldsymbol\mu}\,{\bf b}\,)\,
  =\,{\boldsymbol\mu}\,(\,{\bf a}\,{\boldsymbol\mu})\,{\bf b}\,
  =\,{\boldsymbol\mu}\,(\,{\boldsymbol\mu}\,{\bf a})\,{\bf b} \notag \\
&=(\,{\boldsymbol\mu}\,{\boldsymbol\mu})({\bf a}\,{\bf b})
  =-\,{\bf a}\,{\bf b}
  =-\,{\bf a}\cdot{\bf b}\,-\,{\bf a}\wedge{\bf b} \notag \\
&=-\,{\bf a}\cdot{\bf b}\,-\,{\boldsymbol\mu}\cdot({\bf a}\times{\bf b})
\cong\pm\,1\in S^3,\label{bi-product}
\end{align}
where the definition ${{\boldsymbol\mu}\wedge{\bf x}=\,0}$, associativity of (\ref{fun-cliff}),
and duality relation ${{\bf a}\wedge{\bf b}={\boldsymbol\mu}\,({\bf a}\times{\bf b})}$ have
been used. The last result can be proven also by brute force using (\ref{tri-mu}) and the
Clifford analogues of the triple product identities. Unlike the Pauli identity
(\ref{nowobserve}), the above result is simply a classical relation among various vector
products.

Our next task is to evaluate analogues of the integrals (\ref{class}) and (\ref{classprob}) for
the local functions (\ref{bell-c2}). For this purpose the tool we shall be using is that of a
{\it directed measure} ${d{\boldsymbol\rho}({\boldsymbol\mu})}$, defined on a smooth orientable
{\it vector} manifold ${{\cal V}_3}$ \cite{Clifford}. In other words, we shall be using a
directed measure on a manifold whose ``points'' are vectors in ${{\mathbb E}_3}$, obeying
the Clifford product (\ref{fun-cliff}). It is a matter of indifference relative to which
pseudoscalar we choose to evaluate our local integrals, as long as its orientation is
single-valued and continuous. An obvious choice is ${I}$, which we have defined using a
right-handed frame ${\{{\bf e}_j\}}$ of orthonormal vectors. The directed measure on the
manifold ${{\cal V}_3}$ can then be written as ${d{\boldsymbol\rho}({\boldsymbol\mu})}$ ${=}$
${I\,|d{\boldsymbol\rho}({\boldsymbol\mu})|}$, where ${|d{\boldsymbol\rho}({\boldsymbol\mu})|}$
is a scalar measure of the Riemann integration. A directed integral is thus an oriented
Riemann integral, with the orientation determined by the volume element ${I}$ (which is constant
for us, since ${{\mathbb E}_3}$ happens to be flat). In general, however, since the Clifford
product of any two functions on ${{\cal V}_3}$ can be {\it non-commutative}, the result of
a directed integration may depend on the ordering of its factors.

Using these tools, the isotropically weighted averages of the two outcomes
${{\boldsymbol\mu}\cdot{\bf a}}$ and ${{\boldsymbol\mu}\cdot{\bf b}}$, analogous
to the expectation values (\ref{class}), can be easily worked out, giving
\begin{align}
{\cal E}_{c.v.}({\bf n})=\int_{{\cal V}_3}
&{\boldsymbol\mu}\cdot{\bf n}\;\,d{\boldsymbol\rho}({\boldsymbol\mu})\,
  =\,I^2\!\!\int_{{\cal V}_3}{\bf n}\;sign(\boldsymbol\mu)\;
   |d{\boldsymbol\rho}({\boldsymbol\mu})| \notag \\
&=\,-\,\frac{1}{2}\,{\bf n}\,+\,\frac{1}{2}\,{\bf n}\,=\,0\,, \label{cliff}
\end{align}
where ${{\bf n}={\bf a}}$ or ${\bf b}$, the subscript ${c.v.}$ stands for Clifford variables,
and ${{\boldsymbol\rho}(\boldsymbol\mu)}$ is assumed to be normalized on ${{\cal V}_3}$. 
Similarly, the joint expectation value of the two outcomes in the manifestly local
form (\ref{prob}) works out to be
\begin{align}
{\cal E}_{c.v.}({\bf a},\,{\bf b})
&=\int_{{\cal V}_3}
(\,{\boldsymbol\mu}\cdot{\bf a}\,)
(\,{\boldsymbol\mu}\cdot{\bf b}\,)\;\,d{\boldsymbol\rho}({\boldsymbol\mu}) \notag \\
&=\,-\,{\bf a}\cdot{\bf b}\,-\int_{{\cal V}_3}
{\boldsymbol\mu}\cdot({\bf a}\times{\bf b})\;\,d{\boldsymbol\rho}({\boldsymbol\mu}) \notag \\
&=\,-\,{\bf a}\cdot{\bf b}\,+\,0\,,\label{derive}
\end{align}
where we have used the results (\ref{bi-product}) and (\ref{cliff}).

It is crucial to note here that in our derivation above the measure
${d{\boldsymbol\rho}({\boldsymbol\mu})}$ remains independent of the detector settings
${\bf a}$ and ${\bf b}$, chosen ``freely'' at a later time. Nor have we at any time let
the local variables ${A_{\bf a}({\boldsymbol\mu})}$ and ${B_{\bf b}({\boldsymbol\mu})}$
depend on the remote settings ${\bf b}$ and ${\bf a}$, respectively \cite{Bell-La}.

The above result is of course exactly what is predicted by quantum mechanics. Thus, contrary
to Bell's claim, a local realistic model can indeed be constructed to exactly reproduce
quantum mechanical correlations (\ref{twoobserve}), {\it without necessitating remote
contextuality or backward causation}. This fact immediately raises a question: what has
gone wrong with Bell's proof of his theorem? The answer to this question is not difficult to
discern. In Bell's proof there is a tacit assumption that alternative functions such as
${A_{\bf a}(\lambda)}$ and ${A_{\bf a'}(\lambda)}$ always commute with each other. This,
however, is not true for the Clifford algebra valued functions of the multivector variables
${\boldsymbol\xi}$, satisfying the non-commutative product relations defined in
(\ref{fun-cliff}).

To appreciate this explicitly, let us reconsider the much studied CHSH string of
expectation values \cite{Clauser}\cite{Shimony-2004}\cite{bounds}:
\begin{equation}
{\cal E}({\bf a},\,{\bf b})\,+\,{\cal E}({\bf a},\,{\bf b'})\,+\,
{\cal E}({\bf a'},\,{\bf b})\,-\,{\cal E}({\bf a'},\,{\bf b'}).
\label{CHSH-op}
\end{equation}
For a generic multivector ${\boldsymbol\xi}$, this can be rewritten as
\begin{equation}
\int_{{\cal V}_3}{\cal F}_{c.v.}(\boldsymbol\xi)\;\,
d{\boldsymbol\rho}(\boldsymbol\xi)\,, \label{probint}
\end{equation}
with the local realistic function ${{\cal F}_{c.v.}(\boldsymbol\xi)}$ defined as
\begin{equation}
\begin{split}
{\cal F}_{c.v.}(\boldsymbol\xi):=A_{\bf a}(\boldsymbol\xi)&
\left\{\,B_{\bf b}(\boldsymbol\xi)+B_{\bf b'}(\boldsymbol\xi)\,\right\} \\
&+A_{\bf a'}(\boldsymbol\xi)\left\{\,B_{\bf b}(\boldsymbol\xi)
-B_{\bf b'}(\boldsymbol\xi)\,\right\}.
\end{split}
\end{equation}
If we now use the fact that, by definition, the functions ${A^2_{\bf a}(\boldsymbol\xi)}$,
${A^2_{\bf a'}(\boldsymbol\xi)}$, ${B^2_{\bf b}(\boldsymbol\xi)}$, and
${B^2_{\bf b'}(\boldsymbol\xi)}$ are all equal to ${\pm}$ unity, then the square of
the function ${{\cal F}_{c.v.}(\boldsymbol\xi)}$ simplifies to
\begin{equation}
{\cal F}_{c.v.}^2(\boldsymbol\xi)\,=\,4\,+\,\left[\,A_{\bf a}(\boldsymbol\xi),\,
A_{\bf a'}(\boldsymbol\xi)\,\right]\left[\,B_{\bf b'}(\boldsymbol\xi),\,
B_{\bf b}(\boldsymbol\xi)\,\right]\!,\label{20/20}
\end{equation}
provided we assume that both of the ${A}$'s commute with both of the ${B}$'s, and vice versa:
\begin{equation}
\left[\,A_{\bf n}(\boldsymbol\xi),\,B_{\bf n'}(\boldsymbol\xi)\,\right]\,=\,0\,,
\;\;\;\forall\;\,{\bf n}\;\,{\rm and}\;\,{\bf n'}.
\label{com}
\end{equation}
In quantum field theory the operator analogue of the last relation---which would state
that the operators acting on different subsystems should commute if the subsystems happen
to be spacelike separated---follows from the usual assumption of ``local commutativity''
\cite{Bell-La}. However, in our equation above the commutation relation is between two
ordinary functions, with a sharper geometrical meaning. It would hold whenever the
Clifford product between the two functions happens to be symmetric \cite{Clifford}, at
least after average \cite{Bell-La}, as in the result (\ref{derive}) of our local model.

Returning to the square of the function ${{\cal F}_{c.v.}\!(\boldsymbol\xi)}$ in
(\ref{20/20}), let us first note that if either of the two commutators in it were to
vanish identically, then we would be led to the standard CHSH inequality with bounds
${\pm\,2}$ \cite{Clauser}. However, neither of the two commutators can vanish in
general, because of the dependence of the local functions ${A_{\bf a}(\boldsymbol\xi)}$
and ${B_{\bf b}(\boldsymbol\xi)}$ on the Clifford algebra valued variables. One can
easily see an explicit instance of this by once again using our local model, but now
setting ${A_{\bf a}(\boldsymbol\mu)=I\,{\bf e}_x}$
and ${A_{\bf a'}(\boldsymbol\mu)=-\,I\,{\bf e}_y}$. The relations (\ref{c-algebra}) then
at once leads to the non-vanishing of the commutator
${\left[\,A_{\bf a}(\boldsymbol\mu),\,A_{\bf a'}(\boldsymbol\mu)\,\right]}$. What is more,
even when both commutators individually vanish upon average, their Clifford product within
(\ref{probint}) before and after the average may not vanish in general,
which can again be checked by simple examples. Once these facts are appreciated, it is
easy to establish that
\begin{equation}
{\cal F}_{c.v.}^2(\boldsymbol\xi)\,\leqslant\,4\,+\,2\times 2\,=\,8\,,\label{8}
\end{equation}
since each of the two commutators in (\ref{20/20}) can reach a maximum
value of ${+\,2}$ (which follows from the fact that each product such as
${A_{\bf a}(\boldsymbol\xi)A_{\bf a'}(\boldsymbol\xi)}$ can equal
to either ${+\,1}$ or ${-\,1}$). Consequently, we arrive at the inequality
\begin{equation}
|{\cal F}_{c.v.}(\boldsymbol\xi)|\,\leqslant\,2\sqrt{2}\,.
\end{equation}
Since this inequality holds for all values of ${\boldsymbol\xi}$, using (\ref{probint})
we finally arrive at the violation of CHSH inequality: 
\begin{equation}
\left|\,{\cal E}({\bf a},\,{\bf b})\,+\,{\cal E}({\bf a},\,{\bf b'})\,+\,
{\cal E}({\bf a'},\,{\bf b})\,-\,{\cal E}({\bf a'},\,{\bf b'})\,\right|\,\leqslant\,2\sqrt{2}\,.
\notag \label{My-CHSH}
\end{equation}
This is of course exactly the same result as that obtained within standard quantum mechanics
\cite{bounds}. The difference, however, is that the above inequality is obtained within an
entirely classical, local realistic framework of Bell.

It is worth noting here that, although for definiteness we have used the language of spin
for our analysis above, it can be easily extended to any two-state system. Hence we can
conclude that Bell inequalities must be violated, with precisely the same characteristics
as they are indeed violated in experiments, not only by quantum mechanics, but by {\it any}
theory that correctly implements the algebra of orthogonal directions in the physical space,
namely the Clifford algebra ${{Cl}_{3,0}}$ of 3D space within nonrelativistic domain, or
the Clifford algebra ${{Cl}_{1,3}}$ of spacetime within relativistic domain. Indeed, it is
clear from our analysis that the often cited Tsirel'son bound \cite{bounds} simply reflects
the algebraic properties of the physical space, and hence should not be taken as
characterizing a purely quantum mechanical feature of the EPR-Bohm correlations.

Finally, what can we say about some of the variants of Bell's theorem, such as the
Greenberger-Horne-Zeilinger, or Hardy's variant \cite{GHSZ}? Can our analysis be extended
to these theorems which do not involve any inequalities? No attempt has been made
here to address this question. We believe, however, that such an extension is possible.

I am grateful to Lucien Hardy and other members of the Perimeter Institute for their
hospitality and support.

\end{document}